\documentclass[doublecol]{epl2}
\usepackage{amsmath}
\usepackage{amssymb}
\usepackage{bm}

\title{Dynamical magnetoelectric effects induced by the Dzyaloshinskii-Moriya interaction in multiferroics}
\shorttitle{Magnetoelectric effect in multiferroics}

\author{Chenglong Jia and  Jamal Berakdar}
\shortauthor{ Jia {\it et al.} }

\institute{
  \inst{} Institut f\"ur Physik, Martin-Luther Universit\"at Halle-Wittenberg, Heinrich-Damerow-Str.4
06120 Halle (Saale), Germany\\
}

\abstract{We study the dynamical interplay between ferroelectricity and  magnetism in a multiferroic with a helical magnetic order. We show that the dynamical exchange-striction induces a biquadratic interaction between the
spins and transverse phonons
resulting in quantum fluctuations of the spontaneous ferroelectric polarization
  $\mathbf{P}$
  in the ferroelectric phase. The hybridization between the spin wave and the fluctuation of the electric polarization leads to  low-lying  transverse phonon modes. Those are perpendicular  to   $\mathbf{P}$ and to the helical spins at small wave vector  but then turn parallel to $\mathbf{P}$ at a wave vector close to the magnetic modulation vector. For helical magnetic structure, the spin chirality which determines the direction of $\mathbf{P}$, also possesses a long-range order. Due to the  dynamical Dzyaloshiskii-Moriya interaction, the spin-chirality is strongly coupled to the spin fluctuation which implies an on-site inversion of the spin-chirality  in the ordered spin-1/2 system and results in a finite   scattering intensity of polarized neutrons from a cycloidal helimagnet.}
\begin{document}
\maketitle
\section{Introduction}
Multiferroic compounds in which the electric and the magnetic order coexist are in the focus of current
 research.  Of a particular interest is the possibility of controlling the direction of the spontaneous electric polarization  by a magnetic field \cite{Multiferroics}. Beside this  technological relevance
 of such a strong interplay between the magnetic and electric order parameters it is
  also  fundamentally  interesting to understand how such a coupling comes about and what is
 the microscopic mechanism behind the magnetoelectric (ME) coupling  in multiferroics.

Among the family of multiferroics, two different type of manganites, RMnO$_3$ (R= Gd, Tb, Dy) \cite{RMnO3} and RMn$_{2}$O$_5$ ($R$= rare earth, Tb,Y, Bi) \cite{RMn2O5} play a special role as they exhibit  different microscopic ME coupling mechanisms: In the perovskite multiferroic RMnO$_3$, it's shown experimentally that the onset of helical magnetic order induces spontaneous ferroelectric (FE) polarization, which can be well described by the so-called spin-current model \cite{KNB}. In these compounds, the spin-orbit coupling within the $d(p)$-orbitals of magnetic(oxygen) ions produces an electric polarization of the form \cite{Jonh}, $\mathbf{P} \sim \mathbf{S}_i \times \mathbf{S}_j$.
Non-collinearity of the spins $\mathbf{S}_j$ (at sites $j$) is strictly required in the spin-current model. In the main FE phase of RMn$_{2}$O$_5$ the electric dipole moments are directed along the $b$-axis, the
 spins however are almost collinear in the $ab$ plane which indicates  that  a spin-orbit-driven mechanism can not be the primary source for FE in RMn$_{2}$O$_5$. As an alternative explanation, the (super)exchange-striction \cite{Striction,YMn2O5} is believed to be the  origin of ferroelectricity in RMn$_{2}$O$_5$, $\mathbf{P} \sim \mathbf{S}_i \cdot \mathbf{S}_j$. On the other hand, inspecting carefully the dynamical properties of the multiferroics, we find both, the antisymmetric Dzyaloshiskii-Moriya (DM) interaction and the symmetric magnetostriction play an essential role and need to be taken into account.

Based on the spin-current model, the dynamical properties of DM interaction were studied in Ref.\cite{d-DM,d-GL,GdMnO3}. A collective ME excited mode, so-called electromagnon, was theoretically predicted. Moreover, it was found  \cite{d-DM} that this  new low-lying mode is perpendicular to both the spontaneous polarization and the helical wave vector. It corresponds to a rotation of the spin plane with respect to the axis of the helical wave vector, the  rotation frequency is $\sqrt{SJD}$, where S is the spin value, J is the exchange coupling, and D is the magnetic anisotropy. Electromagnons have been detected in RMnO$_3$ \cite{ME-RMnO3} and Eu$_{0.75}$Y$_{0.25}$MnO$_3$ \cite{EuYMnO3-1}, seemingly consistent with the theoretical analysis. However, a detailed  study of  the terahertz spectrum of Eu$_{1-x}$Y$_x$MnO$_3$ \cite{EuYMnO3-2} revealed that  infrared-absorption along the spontaneous polarization direction is also observable, which is not explained by
theory.

 In this paper we show that the dynamical exchange striction intrinsically generates bi-quadratic coupling between the spin and the transverse acoustic(TA) phonon, $\sim (\mathbf{u}_i^{\perp} - \mathbf{u}_j^{\perp})^2 (\mathbf{S}_i \cdot \mathbf{S}_j)$ where $\mathbf{u}_j^{\perp}$ is a transverse displacement at site $j$ . This dynamical coupling does not contribute any additional static electric polarization but induces the fluctuation of the electric dipole moment due to the low frequency excitation modes of TA phonon. One thus has a mode mixing behavior and the polarization correlation function follows the soft magnetic behavior of the system  parallel to the uniform electric polarization $\mathbf{P}$. Moreover, in $S=1/2$ multiferroics, the spin-fluctuation is accompanied with an inversion of the local spin and consequently with an inversion of the on-site electric dipole moment   according to the spin-current model. The hybridization between phonons and spins results also in a large quantum fluctuation of the spin-chirality which allows for a finite differential scattering intensity of polarized neutrons from a cycloidal magnet LiCu$_2$O$_2$ \cite{LiCu2O2}.
\section{Dynamical exchange-striction}
We consider a one-dimensional spin chain along the $z$-axis with a frustrated spin interaction. When the temperature is lowered a spiral magnetic structure is realized \cite{LiCu2O2,NaCu2O2}. For such a  helically ordered magnetic phase  the spin-current model predicts a uniform electric polarization perpendicular to the spin chain. The macroscopic electrical polarization  $\mathbf{P}$ is induced by the condensation of the transverse optical(TO) phonons, $\mathbf{P}= - e \mathbf{u}_0$. An effective model can be introduced to describe the spin-phonon coupling \cite{d-DM} as follows

\begin{eqnarray}
&& H = H_s + H_{DM} + H_p \\
&& H_s = \sum_{\langle ij \rangle_{nn}} J_1(r_i-r_j) S_i \cdot S_j \nonumber\\
 && \quad \quad + \sum_{\langle lm \rangle_{nnn} } J_2(r_l-r_m) S_l \cdot S_m \nonumber \\
&& H_{DM} = - \lambda \sum_i \mathbf{u}_i \cdot [\hat{e}_z \times (S_i \times S_{i+1})] \nonumber \\
&& H_p = \frac{k}{2} \sum_i \mathbf{u}_i^2 + \frac{1}{2M} \sum_i \mathbf{P}_i^2 \nonumber
\end{eqnarray}
where the notation $\langle ij \rangle_{nn}$ indicates  nearest-neighboring (nn)  $i$ and $j$, and $\langle lm \rangle_{nnn} $ corresponds the next-nearest-neighboring (nnn) $l$ and $m$. The competition between the nn ferromagnetic interaction ($J_1<0$) and the nnn antiferromagnetic interaction ($J_2>0$) leads to magnetic frustration and  to a spiral spin ordering with the wave vector $\cos Q = -J_1/4J_2$ \cite{NaCu2O2}. $H_p$ is an optical phonon model. The spin-phonon interaction $H_{DM}$ originates from a spin-orbital (DM) coupling and breaks the inversion symmetry along the chain.
Minimizing the energy yields the condition of the atomic displacement and the local spin-configuration
\begin{eqnarray}
&& \mathbf{u}_{i} =\frac{\lambda}{k}\hat{e}_{z}\times(
\mathbf{S}_{i}\times \mathbf{S}_{i+1})
\label{eq::displacement}
\end{eqnarray}
Particularly, if the $zx$ helical spins along the chain, i.e. $\mathbf{S}_i = S (\sin iQ , 0, \cos iQ)$,
Eq.(\ref{eq::displacement}) leads to a macroscopic uniform lattice displacements along the $x$ direction $\mathbf{u}_0^x = - \frac{\lambda S^2}{k} \sin Q \hat{e}_x$.

In the helical spin-ordering phase 
 $\mathbf{u}_x$ can't be softened through the hybridization between the TO phonons and 
the  magnons because $k/M \gg JS$. The spontaneous FE polarization $\mathbf{P}_x$ is frozen  at $-e\mathbf{u}_0^x$. The fluctuation $\delta P_x$  can therefore  be neglected \cite{d-DM}. However,  accounting for the superexchange striction,  TA phonon mode emerges. As well-known,  TA phonons possess a low frequency mode at the long wavelength, $\omega_{TA}^2(q) \propto  q^2$ which gives rise to the fluctuation of the FE polarization. Such polarization fluctuations are hybridized with the spin bosons and soften thus the transverse phonon behavior.

Existing experimental data suggests that the exchange energy $J$ falls off as a power law  with the separation of the magnetic ions
\begin{eqnarray}
J_{1,2}(r_i-r_j) = J_{1,2}|(R_i^z + \mathbf{u}_i) - (R_j^z + \mathbf{u}_j)|^{-\gamma_{1,2}},
\end{eqnarray}
where $\gamma$  is in the range of $6-14$ \cite{Book}. $R_i^z$ is the bare value of the position of the atom at site $i$, and $|R_i^z - R_j^z|$ determines the lattice constant $a$ (set here to 1). $\mathbf{u}_i$ is the displacements of site i. Generally, $|\mathbf{u}_i|$ is small and does not destroy the lattice structure, ($|\mathbf{u}_i|/a \sim 10^{-3}$). We inspect the dominant term for $J$ in the following two case.

\emph{Longitudinal phonons}. When the atoms are displaced along the chain one finds
\begin{equation}
J_{1,2}(r_i-r_j) =
J_{1,2}[1-{\gamma_{1,2}} ~ \hat{e}_{ij} \cdot (\mathbf{u}_i^z -\mathbf{u}_j^z)]
\end{equation}
with $\hat{e}_{ij}$ being the unit vector connecting two sites $i$ and$j$. A trillinear coupling between the phonon and spin is induced. One can easily check that $\langle \mathbf{u}_i^z \rangle = 0$ because of the local rotational symmetry of the spin-spin correlation $\langle \mathbf{S}_i \cdot \mathbf{S}_j \rangle$ in the helically ordered phase. Since $\mathbf{u}_i^z$ is not involved in the DM interaction we do not  consider it in the following discussion.

\emph{Transverse phonons}. For atomic displacements   perpendicular to the chain, $\mathbf{u}_i^{\perp} \cdot \hat{e}_z=0$ we find for $J_{1,2}(r_i-r_j)$

\begin{equation}
J_{1,2}(r_i-r_j) \approx
J_{1,2}[1-{\gamma_{1,2} \over 2}(\mathbf{u}_i^{\perp} -\mathbf{u}_j^{\perp})^2]
\label{ES}
\end{equation}
which gives a TA phonon mode coupled to the spins with the bi-quadratic interaction
%
$- \gamma_{1,2} J_{1,2} (\mathbf{u}_i^{\perp} -\mathbf{u}_j^{\perp})^2 (\mathbf{S}_i \cdot \mathbf{S}_j)$.
%
The effective spring constant for the TA phonon is $k_{TA} = \gamma J \langle \mathbf{S}_i \cdot \mathbf{S}_j \rangle$. Furthermore, because of the negative $ J_1 \langle \mathbf{S}_i \cdot \mathbf{S}_j \rangle + J_2 \langle \mathbf{S}_l \cdot \mathbf{S}_m \rangle $ the dynamical exchange-striction will harden the frequency of the transverse phonon mode as
\begin{eqnarray}
\tilde{\omega}_0^2 = \omega_0^2 ( 1 + |F_s|)
\end{eqnarray}
where $\omega_0^2 =k/M$ and $F_s = [ -\gamma_1 J_1 \langle \mathbf{S}_i \cdot \mathbf{S}_j \rangle- \gamma_2 J_2 \langle \mathbf{S}_l \cdot \mathbf{S}_m \rangle ] /k$. As the temperature decreases below the transition temperature for magnetic
order  $T_N$
  the spin-spin correlation function $\langle \mathbf{S}_i \cdot \mathbf{S}_j \rangle$ increases and so does the phonon frequency. Lowering further the temperature to the FE transition temperature $T_{FE}$, an additional frequency hardening occurs due to the dynamical DM interaction \cite{d-DM,GdMnO3}. The complete scenario is thus that
 the phonon frequency hardens at  two onsets at $T_N$ and $T_{FE}$,  a conclusion
 consistent with the experimental observation for Eu$_0.75$Y$_0.25$MnO$_3$ \cite{EuYMnO3-1}.  Assuming $k \sim 1 eV / {\AA}^2$  and $J S^2 \sim 10 meV$ \cite{d-DM} the frequency  hardening can be estimated to be $\delta \omega / \omega_0 \approx 1\%$, which is in good agreement with the experimental data \cite{EuYMnO3-1}.
Phenomenologically, the exchange-striction suggests that in a  Ginzburg-Landau (GL) theory for the coupling between the spin $S$ and the transverse electric dipole $P_{\perp}$  terms of the form $-\alpha S^2 P_{\perp}^2$ appear. As a consequence, $P_{\perp}$ and $S$ condense at the same temperature due to the strong spin-lattice coupling  $\alpha \sim J$ and the two transition temperatures merge, a conclusion which is in line   with the experimental observations in YMnO$_3$ \cite{YMnO3}. There the electric dipole moment $\Delta P_z$, which is along the $z$ direction obeys  the same temperature dependence as the magnetic moment that is aligned in the $ab(xy)$ plane with $120^{\circ}$ structure below $80K$.

\section{Electromagnon}
We split the atomic displacements into two parts: (i) the statical part $\mathbf{u}_i = (u_0^x, 0, 0)$ driven by the DM interaction, and (ii) the dynamical part $\delta \mathbf{u}_i = (- \delta u_i^x, \delta u_i^y, 0)$ induced by the exchange-striction.
As the softness of the system is due to the magnetic part  we  concentrate at first on the spin excitations.
For the $zx$ helical spins, it is convenient to rotate the
spins locally (at each site) along its classical direction ($\tilde{S}_i^z$)
\begin{eqnarray}
& S_{i}^{x}=\tilde {S}_{i}^{x}\cos iQ + \tilde{S}_{i}^{z}\sin iQ  \\
& S_{i}^{y}=\tilde{S}_{i}^{y} \\
& S_{i}^{z}=-\tilde{S}_{i}^{x}\sin
iQ +\tilde{S}_{i}^{z}\cos iQ .
\end{eqnarray}
Disregarding the high-order terms of the interplay between the spins and the dynamical part of lattice displacements, i.e. using the standard linear-spin-wave approximation, we have

\begin{eqnarray}
H &=& E_0 + \sum_q A(q) \tilde{S}_{q}^{-}\tilde{S}_{q}^{+} + B(q) ( \tilde{S}_{q}^{-}\tilde{S}_{\bar{q}}^{-} + \tilde{S}_{q}^{+}\tilde{S}_{\bar{q}}^{+}), \nonumber \\
\end{eqnarray}
where $E_0 = N [J_1 S^2 \cos Q + J_2 S^2 \cos 2Q - \frac{k}{2} |u_0^x|^2]$, and

\begin{eqnarray}
A(q) &=& -J_1[\cos Q + \frac{1}{2}(1+ \cos Q)\cos q] \nonumber \\ &-& J_2[\cos 2Q + \frac{1}{2}(1+ \cos 2Q)\cos 2q] \nonumber \\ &+& \frac{\lambda^2 S^2 \sin^2 Q }{2k}(2-\cos q) \\
B(q) &=& {J_1 \over 4} (1-\cos Q) \cos q + {J_2 \over 4}(1- \cos 2Q) \cos 2q \nonumber \\ &-& \frac{\lambda^2 S^2  \sin^2 Q }{4k}\cos q
\end{eqnarray}
$H$ can be easily diagonalized by a Baguliubov transformations. The energy dispersion of the spin-excitation reads

\begin{equation}
\omega_s (q) = [A(q)^2-(2B(q))^2]^{1/2}.
\end{equation}
%
The effective spin anisotropy introduced by the spin-phonon(DM) interaction results in an energy gap of the spin-wave spectrum  for non-collinear spin ordering, i.e. $\omega_s (q=Q) \neq 0$ if $Q \neq 0~ or ~ \pi$. One can see in the further discussion that the spin fluctuation with the wave vector $q \approx Q$ are important in connection with  the \emph{magnetic} softening of  the transverse phonons.

Now we turn our attention to the dynamical spin-phonon interaction.
Retaining terms up to the second order in the quantum fluctuation, the spin-current model delivers the following coupling terms
%
\begin{eqnarray}
\tilde{H}_{DM} &=& -\lambda S \cos Q \sum_i \delta u_i^x (\tilde{S}_{i+1}^x - \tilde{S}_i^x) \nonumber \\ &~~~& - \lambda S \sum_i \delta u_i^y (\tilde{S}_i^y \cos Q_{i+1} - \tilde{S}_{i+1}^y \cos Q_i) \nonumber \\
&=& -\lambda S \cos Q \sum_q \delta u_q^x \tilde{S}_q^x ( \cos q -1) \nonumber \\ &~~~& - \lambda S \sum_q \delta u_q^y \tilde{S}_{q\pm Q}^y ( e^{\mp i Q} - e^{i(q \pm Q)}) /2
\label{dME}
\end{eqnarray}
%
$\delta u_q^y$ is hybridized with the spin at $q \pm Q$, but $\delta u_q^x$ is coupled to $\tilde{S}^x$ at $q$. As   expected, $\delta u_q^x$ has the same long wavelength behavior as the magnons. No static displacement  exists
along the  $x$ direction, i.e.  $\delta u_0^x =0$. On the other hand, a uniform lattice deformation along the $y$ direction ($\delta u_0^y \neq 0$) may occur  due the hybridization between the electric polarization and the spin ordering\cite{d-DM}.
After some algebra, we find for the  polarization correlation functions
\begin{eqnarray}
&& \ll \delta u_q^x | \delta u_{\bar{q}}^x \gg = \frac{\omega^2 - \omega_s^2}{M[\omega^4 - \omega^2(\omega_p^2 + \omega_s^2) + \omega_p^2 (\omega_{s}^2-\omega_{sp}^2)]}\nonumber \\
&& \ll \delta u_q^y | \delta u_{\bar{q}}^y \gg = \frac{1}{M[\omega^2 - \omega _p^2 +  \frac{\lambda^2 S^3}{2M}\sum_{q'=q \pm Q}G_{s}(q')]}
\nonumber\end{eqnarray}
where $\omega_p = \sqrt{\omega_0^2 + \omega_{TA}^2}$ is the frequency for the transverse phonon, $\omega'_{sp}(q)= [2(A(q)-2B(q))(\lambda ^2 S^3 \cos^2 Q (1- \cos q))/k']^{1/2}$, and $G_{s}(q \pm Q) = (A(q \pm Q) + 2 B(q \pm Q))(1- \cos (q \pm 2Q)/(\omega^2 - \omega_s ( q \pm Q))$. At small wave vectors, $q \sim 0$ and $\omega_{TA}(0) \sim 0$, the TA phonon is decoupled from spins.
The antisymmetric DM interaction dominates over the spin-phonon coupling. $\delta u_0^y $ is coupled  via $(\tilde{S}_Q^y - \tilde{S}^y_{\bar{Q}})$  to the rotation of the spin plane and the direction of the polarization along the chain.
 Additionally a uniform polarization in the  $y$ direction is induced by the dynamical ME interaction and $\ll \delta u_0^y | \delta u_0^y \gg$ possesses a low frequency behavior. The rotation mode  around the $z$ axis  has $\omega_{-}^{y} \sim \sqrt{JSD}$ if an easy-plane spin anisotropy $D (S_y)^2$ is introduced to the spin system. However, at a wave vector close to the magnetic modulation vector, i.e. $q \sim Q$ and $\omega_{TA}(Q) \neq 0$ both the symmetric and antisymmetric magnetoelectric interaction respond to the fluctuations of the polarization.  Especially, in the direction parallel to the FE polarization  $P_x$, there is a low frequency range around  $\omega_{-}^x \cong \omega_s (Q)$ where $u^x$ couples  resonantly to light even if $D=0$. For finite $D$, assuming $D \gg \lambda^2 S^2/k$  we observe  nearly the same low-frequency behavior of the polarization correlation functions $\omega_{-}^x \approx \sqrt{JSD} \approx \omega_{-}^y$. These conclusions are also  qualitatively  consistent with experiment observations (Fig.8 in Ref.\cite{EuYMnO3-2}).
\section{Spin-flip}
Recently, LiCu$_2$O$_2$ ($S=1/2$) has been found to be ferroelectric in the bc-spiral state at low temperatures \cite{LiCu2O2}.
In contrast to  large spin multiferroics, in spin-1/2 magnet the spin fluctuations may spontaneously reverses  the local spin. According the spin-current model Eq.(\ref{eq::displacement}) the direction of the on site electric dipole moment can also be completely reversed by the spin fluctuations. Large quantum fluctuations of the FE polarization $\delta \mathbf{u}_i^x = - 2 \mathbf{u}_0^x$ is induced by the hybridization between the phonon and the spin.

Defining the vector of spin chirality as the average of the outer product of two adjacent spins $\hat{c}_i = (s_i \times s_{i+1})/|s_i \times s_{i+1}|$, in the RMnO$_3$-type multiferroics the direction of electric polarization is determined by the spin chirality. Reversing the direction of electric polarization does also reverse $\hat{c}_i$. Clearly, the spin chirality $\hat{c}_i$ has only two eigenvalues, $+1$ and $-1$, and possesses long-range \emph{ferromagnetic} order in the FE phase. Thus, $\hat{c}_i$ can be simply treated as the Pauli operator. According to the dynamical exchange-striction Eq.(\ref{ES}) the interaction term
 involving the spin chirality has the structure $\sim - J_c(Q) \hat{c}_i \cdot \hat{c}_{i+1}$. The $x$ component of the spin-chirality operator $\hat{c}_i^x$ acts as a direction reversal operator  which can be traced back to the quantum fluctuation of the FE polarization. Considering the the dynamical DM interaction Eq.(\ref{dME}),  the coupling term  between the spin and the spin-chirality in the spin-1/2 multiferroics is given by
%
$\sum_i \hat{c}_i^x(\hat{s}_{i+1}^x - \hat{s}_i^x) = \sum_i \hat{s}_i^x (\hat{c}_{i-1}^x - \hat{c}_i^x)$,
%
which indicates that  when the spin at site $i$ is flipped, $\hat{s}_i \rightarrow -\hat{s}_i$, the direction of spin-chirality $\hat{c}_i$ and $\hat{c}_{i-1}$  are also reversed, an observation consistent with the spin-current model and with the definition of the spin-chirality.

For the one-dimensional spin-1/2 chain, the quantum model predicts a gapped spin-liquid state in the range of the frustration exchange parameters in LiCu$_2$O$_2$. The very existence of the magnetic helix state suggests that the quantum fluctuations is significantly suppressed and  the spins tend to recover  a semiclassical behavior. In the ground state of the spin system  all spins point along their corresponding classical directions as in NaCu$_2$O$_2$, where a $J_1-J_2$ spin model provides a good description of the helix state \cite{NaCu2O2}. So the spin interaction can be simply given as $-J_s(Q) \hat{s}_i \cdot \hat{s}_j$ where $Q$ is taken as the pitch angle along the chain.

Based on the above arguments an effective model that describes the interplay between the helical spin and spin-chirality has the form
\begin{eqnarray}
H_{sc} = -\sum_{i,j} (J_s \hat{s}_i \cdot \hat{s}_j + J_{c} \hat{c}_i \cdot \hat{c}_j) - \gamma \sum_i \hat{s}_i^x (\hat{c}_{i-1}^x - \hat{c}_i^x).
\end{eqnarray}
  The Hilbert space can be considered as the tensor product space 
$
|i \rangle \rightarrow |s_i^z \rangle _s \otimes | c_i^z \rangle _c.
$
The ground state of $H_{sc}$ possesses  the ferromagnetic order both for $\hat{s}$ and $\hat{c}$, i.e. $|g.s.\rangle = |FM \rangle_s \otimes |FM \rangle_c$. Now let us consider the effect of the quantum fluctuation. If the spin at site $i$ is flipped we have $|s_i \rangle = \hat{s}_i^x |g.s. \rangle = |\bar{s}_i^z \rangle _s \otimes |FM \rangle_c $ which is the ground state with the spin at site $i$ being flipped. Noting that
%
 $\hat{s}_i^x |g.s. \rangle = |\bar{s}_i^z \rangle _s \otimes |FM \rangle_c, ~~ (\hat{s}_i^x)^2 |g.s. \rangle =  |g.s. \rangle,\;\;
 \hat{c}_i^x |g.s. \rangle = |FM \rangle_s \otimes |\bar{c}_i^z \rangle_c, ~~ (\hat{c}_i^x)^2 |g.s. \rangle =  |g.s. \rangle,
$
%
if we apply  $H_{sc}$ to the state
$\left( \begin{array} {cc}| s_q \rangle \\  |c_q \rangle \end{array} \right)$ we find
\begin{eqnarray}
 H_{sc}  \left( \begin{array}
{cc}| s_q \rangle \\  |c_q \rangle
\end{array} \right ) = \left[ E_0(q) + E_{sp} \left(
\begin{array}{cc}
\cos \theta & \sin \theta \\
\sin \theta & -\cos \theta
\end{array} \right) \right] \left( \begin{array}
{c} | s_q \rangle \\ | c_q \rangle
\end{array} \right ) \nonumber
\end{eqnarray}
where the state $|s_q \rangle$ ($|c_q \rangle$) is essentially a flipped spin (spin-chirality) delocalized across all the lattice.  $E_0 = -NJ_s s^2 -N J_c c^2 + [J_s(q)+J_c (q)]/2$, $J_s(q) = 2s J_s (1- \cos q)$, $J_c(q) = 2c J_c (1- \cos q)$, and $E_{sp} = \sqrt{ [ (J_s (q) - J_c (q))/2]^2 + \gamma (p)^2}$, $\gamma (p) = \gamma (1- \cos q)$, $\cos\theta = ( J_s(p) - J_c (p) )/ 2E_{sp}$, and $\sin\theta = \gamma (p) /E_{sp}$. Applying the rotation
%
$ \left(\begin{array}{c}  |\tilde{c}_q \rangle  \\|\tilde{s}_q \rangle \end{array} \right)
=\left(  \begin{array}{cc} -\sin \theta/2 & \cos\theta/2 \\
\cos \theta/2 & \sin\theta/2
\end{array} \right) \left(\begin{array}{c} | s_q \rangle \\ |c_q \rangle \end{array}\right)$
%
the Hamiltonian is brought in the diagonal form
\begin{eqnarray}
H_{sc}= \sum_q (E_0(q) + E_{sp} ) | \tilde{s}_q \rangle + \sum_q (E_0(q) - E_{sp} ) | \tilde{c}_q \rangle .\nonumber
\end{eqnarray}
Due to the spin-phonon coupling the spin and spin-chirality excitations are mixed. Two separated channels
 are identified: the spin-channel $ |\tilde{s}_q \rangle$ and the phonon-channel $| \tilde{c}_q \rangle$. In each channel, we have
\begin{equation}
\langle \hat{s} \rangle + \langle \hat{c} \rangle =1.
\label{E-value}
\end{equation}
Generally, the expected value of $\hat{c}$ is less then one due  hybridization with  spin excitations. A non-unitary $\hat{c}$  is the origin for a finite  scattering intensity of polarized neutrons: For the cycloidal helimagnet, we have \cite{Tokura}
\begin{eqnarray}
\langle \hat{c} \rangle = \frac{I_{on} - I_{off}}{I_{on} + I_{off}}
\end{eqnarray}
where $I_{on}$($I_{off}$) is the reflection intensity of polarized neutrons parallel(antiparallel) to the scattering vector.
On the basis of the experimental data for LiCu$_2$O$_2$ \cite{LiCu2O2} we infer  $\langle \hat{c} \rangle \approx 0.3$. On the other hand, the magnitude of the ordered moment per magnetic copper site is $0.56 \mu_B$ \cite{NaCu2O2}. Together with the typical g-factor for Cu$^{2+}$ in a square-planar geometry ($g \approx 2$) 
from Eq.(\ref{E-value}) we conclude  $\langle \hat{c} \rangle = 0.44$  which is consistent with the previous estimated value.

Summarizing,  both the (symmetric) exchange-striction and (antisymmetric) DM interaction affect dynamically the magnetoelectric coupling  in multiferroics.
At a small wave vector, the DM interaction determines  the low-frequency behavior of the phonons.
For a wave vector close to that of the magnetically modulated structure, the exchange striction induces fluctuations in the FE polarization, and additional low-lying mode parallel to the FE polarization  emerges. For spin-1/2 multiferroics, the effect of the quantum fluctuation is particularly   large. The local polarization can be completely reversed by the spin fluctuation, and so does the direction of the on site spin-chirality.
These findings are in line with experimental observations. \\
{\it Acknowledgement:} This work is supported by the German Science Foundation DFG through  SFB762 -B7- {\it functionality of oxide interfaces}.

\end{document}